\documentclass[twocolumn,english,prl,showpacs]{revtex4}
\usepackage[T1]{fontenc}
\usepackage[latin9]{inputenc}
\usepackage{color}
\usepackage{amsmath}
\usepackage{amssymb}
\usepackage{graphicx}
\usepackage{esint}

\makeatletter

\providecommand{\tabularnewline}{\\}

\@ifundefined{textcolor}{}
{%
 \definecolor{BLACK}{gray}{0}
 \definecolor{WHITE}{gray}{1}
 \definecolor{RED}{rgb}{1,0,0}
 \definecolor{GREEN}{rgb}{0,1,0}
 \definecolor{BLUE}{rgb}{0,0,1}
 \definecolor{CYAN}{cmyk}{1,0,0,0}
 \definecolor{MAGENTA}{cmyk}{0,1,0,0}
 \definecolor{YELLOW}{cmyk}{0,0,1,0}
}

\makeatother

\usepackage{babel}
\begin{document}

\title{Field-reversed bubble in deep plasma channels for high quality electron
acceleration }

\author{A. Pukhov$^{1}$, O. Jansen$^{1}$, T.Tueckmantel$^{1}$, J. Thomas$^{1}$
and I. Yu. Kostyukov$^{2,3}$}

\affiliation{$^{1}$Institut fuer Theoretische Physik I, Universitaet Duesseldorf,
40225 Germany}

\affiliation{$^{2}${\small{}Lobachevsky National Research University of Nizhni
Novgorod, 603950, Nizhny Novgorod, Russia}}

\affiliation{$^{3}${\small{}Institute of Applied Physics RAS, Nizhny Novgorod
603950, Russia}}
\begin{abstract}
We study hollow plasma channels with smooth boundaries for laser-driven
electron acceleration in the bubble regime. Contrary to the uniform
plasma case, the laser forms no optical shock and no etching at the
front. This increases the effective bubble phase velocity and energy
gain. The longitudinal field has a plateau that allows for mono-energetic
acceleration. We observe as low as $10^{-3}$ r.m.s. relative witness
beam energy uncertainty in each cross-section and 0.3\% total energy
spread. By varying plasma density profile inside a deep channel, the
bubble fields can be adjusted to balance the laser depletion and dephasing
lengths. Bubble scaling laws for the deep channel are derived. Ultra-short
pancake-like laser pulses lead to the highest energies of accelerated
electrons per Joule of laser pulse energy.
\end{abstract}

\pacs{PACS1}

\maketitle
The laser wake field acceleration (LWFA) \cite{esarey} in underdense
plasmas provides an option for high gradient particle acceleration
\cite{Joshi}. Especially efficient is the \textit{bubble regime}
\cite{bubble} when the laser expels background plasma electrons from
the first half of the plasma wave. The advantage of the cavitated
region is the transverse uniformity of its axial accelerating field
\cite{Kost2004} so that the energy gain of the accelerated electrons
is not affected by their lateral motions in the bubble and the bunch
can remain monoenergetic. The quasi-monoenergetic electron bunches
are readily registered in experiments \cite{Malka 2013}. 

\textsl{\emph{Despite various theoretical approaches: the phenomenological
model of the bubble \cite{Kost2004}, the nonlinear theory of blowout
regime \cite{scaling Lu,BlowoutTheory Lu} and the similarity theory
\cite{similarity}, - a self-consistent analytical description of
the bubble regime is still absent.}} 

The bubble theories exist for homogeneous plasmas only. Similarity
theory together with energy conservation arguments resulted in the
``optimal'' scaling laws for homogeneous plasma \cite{scaling PG}.
These were tested in 3d PIC simulations \cite{Jansen2014}. The scalings
assume that the laser energy is converted into the bubble fields first
and then harvested by the electron bunch. The leading similarity parameters
for the bubble in homogeneous plasmas are the $S-$number $S=n_{e}/an_{c}\ll1$
and the pulse aspect ratio $\Pi=c\tau/R\leq1$. Here, $n_{e}$ is
the plasma electron density and $n_{c}=\pi/r_{e}\lambda_{0}^{2}$
is the critical plasma density for a laser pulse with the wavelength
$\lambda_{0}$, $a=eE_{0}/mc\omega_{0}$ is the dimensionless laser
amplitude, $\omega_{0}=2\pi c/\lambda_{0}$, and $r_{e}=e^{2}/mc^{2}$
is the classical electron radius, $\tau$ is the pulse duration and
$R$ is its radius. 

In a uniform plasma, the bubble regime has some drawbacks. First,
the laser energy depletion length is shorter than the electron dephasing
length. This limits the maximum electron energy gain. Second, the
transverse bubble fields acting on the accelerated electron bunch
are strongly focusing. Thus, electrons running forward with the relativistic
factor $\gamma$ oscillate about the bubble axis at the betatron frequency
$\omega_{\beta}=\omega_{p}/\sqrt{2\gamma}$, where $\omega_{p}=\sqrt{4\pi e^{2}n_{e}/m}$
is the background plasma frequency. A relativistic electron can easily
come into betatron resonance with the Doppler-shifted laser that may
result in energy exchange \cite{DLA}. This betatron resonance broadens
the electron bunch energy distribution and deteriorates the beam quality
\cite{Dino Nature,DLA Mori}.

Here, we consider the bubble regime of electron acceleration, but
in a deep plasma channel rather than in a uniform plasma. This allows
for a significant improvement of the acceleration . \emph{We demonstrate
that (i) the effective bubble phase velocity and energy gain increase
in the channel; (ii) the longitudinal field has a plateau that allows
for mono-energetic acceleration; (iii) the focusing force acting on
the accelerated bunch is strongly reduced; (iv) the bubble scaling
laws and the bubble field distribution for the deep channel are derived;
(v) according to the new scaling laws,} \textit{ultra-short pancake-like
laser pulses match the dephasing and depletion length in the channel
and thus lead to the highest energy gains of accelerated electrons
per Joule of laser pulse energy}. In simulations, we observe as low
as $10^{-3}$ r.m.s. relative witness beam energy uncertainty in each
cross-section and 0.3\% total energy spread. The lack of focusing
in the channel eliminates the very possibility of a betatron resonance
and leads to much sharper beam energy distributions.

Normally, plasma channels are used to guide weakly relativistic pulses
over distances much larger than the Rayleigh length $Z_{R}=\pi R^{2}/\lambda_{0}$
. These channels are shallow as a rule, i.e., the relative on-axis
plasma density depletion is slight. Schroeder et al. \cite{Schroeder 2013}
suggested recently the use of nearly hollow plasma channels to provide
independent control over the focusing and accelerating forces. Chiou
et al. \cite{Chiou} revealed that the transverse electric field of
the wake is not monotonic in the hollow channel. However, these papers
were limited only to a quasilinear regime of the plasma wake generation
and a rectangular channel density profile.

A plasma channel is not required for laser guiding in the bubble regime,
since the laser pulse is self-guided by the cavitated region of the
bubble. Thus, a preformed shallow plasma channel is not expected to
change the bubble dynamics significantly. However, as will be seen,
a deep plasma channel i.e., one that is (nearly) empty on-axis, can
strongly modify the bubble fields, the nonlinear laser dynamics, and
the trapping. We are looking for laser-plasma parameters that maximize
energy of the accelerated electron bunch and improve its quality:
mainly, reduce the energy spread.

The laser pulse energy $W_{L}$ is the technologically important characteristics
of a laser pulse. It is limited, e.g. by size of the active crystal
or by compression gratings, etc. We are comparing the energy scalings
for the two cases: the uniform plasma case and the deep channel case.

In uniform plasma, the similarity scalings for electron bunch energy
$\mathcal{E}_{\max}$ from a bubble \cite{scaling PG} can be expressed
in terms of the laser pulse power $P_{L}$, duration $\tau$ and energy
$W_{L}=P_{L}\tau$: 
\begin{equation}
\mathcal{E}_{\max}^{{\rm uniform}}\propto\sqrt{P_{L}e^{2}c}\tau\lambda_{0}^{-1}=\lambda_{0}^{-1}\sqrt{W_{L}e^{2}c\tau}\label{eq:uniform}
\end{equation}
 The scaling of Eq. \eqref{eq:uniform} favors longer laser pulses.
This reflects the mechanism of laser pulse depletion in the standard
bubble. The laser pulse interacts with plasma electrons at the very
front only. The major part of the pulse propagates freely in the cavitated
region. Thus, the pulse tail slowly overtakes its head, where an ``optical
shock'' is formed and the pulse ``etches''. This ``etching''
leads to the pulse shortening \cite{Faure 2005 shortening} and faster
dephasing that in turn lowers the maximum energy gain. The highest
electron energies are achieved with ``spherical'' laser pulses,
whose duration $c\tau$ equals their radius $R$ \cite{Jansen2014}.
Such pulses fill the cavitated region completely and interact with
the accelerated bunch affecting its quality. 

\begin{table}
\begin{tabular}{|c|c|c|c|}
\hline 
$W_{{\rm L}}$, J & 2.2 & 17.6 & 141\tabularnewline
\hline 
$\tau$, fs & 4 & 8 & 16\tabularnewline
\hline 
$R$,$\mu$m & 8 & 16 & 32\tabularnewline
\hline 
$n_{0}$, 1/cm$^{3}$ & $5\cdot10^{18}$ & $1.25\cdot10^{18}$ & $0.31\cdot10^{18}$\tabularnewline
\hline 
$R_{{\rm ch}}$,$\mu$m & 6 & 12 & 24\tabularnewline
\hline 
$l_{i}$, $\mu$m & 0.8 & 1.6 & 3.2\tabularnewline
\hline 
$N_{i}$, pC & 10 & 20 & 40\tabularnewline
\hline 
$\mathcal{E}_{{\rm i}}$, GeV & 0.1 & 0.4 & 1.6\tabularnewline
\hline 
$L_{{\rm A}}$, cm & 1.6 & 12.8 & 100\tabularnewline
\hline 
$\mathcal{E}_{\max}^{{\rm dc}}$, GeV & 1.5 & 6 (6) & 23.5 (24)\tabularnewline
\hline 
\end{tabular}\protect\caption{\label{fig:table}Laser-plasma parameters used in the simulations
and the observed electron energies. Expected energies from the scaling
law are in brackets. All the lasers had $a_{0}=10$; the channel parameter
$\delta=0.3$.}
\end{table}

Also, in the uniform plasma case, the bubble may continuously trap
electrons causing strong beam loading and thus large energy spread.
Also, the accelerating field in the standard bubble is a linear function
of the longitudinal coordinate \cite{Kost2004,scaling Lu}. If one
wants a truly mono-energetic acceleration of externally injected particles,
the witness bunch to be accelerated must be extremely short. Demanding,
e.g. 1\% energy spread, the witness bunch should not occupy more than
1\% of the bubble length. 

\begin{figure}
\includegraphics[width=0.7\columnwidth]{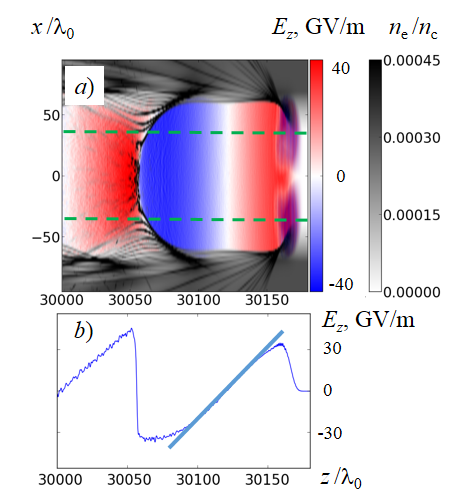}

\protect\caption{\label{fig:bubble} Simulation results for the case $R=32\,\mu$m.
a) The cross-section of the plasma electron density and the longitudinal
bubble field; b) on-axis accelerating field of the bubble, GV/m. Flattening
of $E_{z}$ at the bubble rear part is visible. The broken lines in
frame (a) mark the vacuum part of the plasma channel. The purple rippling
of the $E_{z}$ at the bubble head is caused by the short laser pulse.}
\end{figure}

The use of the deep plasma channel gives us several new control parameters
to improve the acceleration. We consider channels with a zero plasma
density on-axis and some smoothly growing density at the borders.
We have chosen the following parameterization for the plasma radial
profile: $n_{e}=n_{0}\left[\delta\exp(r/R_{{\rm ch}})-1\right]$,
for\emph{ $r\geq r_{0}$ }\textsl{\emph{and }}\emph{$n_{e}=0$ }\textsl{\emph{for}}\emph{
$r<r_{0}$, }\textsl{\emph{where}}\emph{ $r_{0}=R_{ch}\ln(1/\delta)$}.
Such profiles can be produced by ablation from walls of an empty capillary
\cite{ziegler 2009}. The parameters $n_{0},\,\delta,\, R_{{\rm ch}}$
provide enough freedom to adjust fields in the channel while ensuring
a smooth guiding of the relativistically intense laser pulse. In simulations,
we reduced the maximum plasma density far from the channel axis for
numerical stability.

The freedom on the bubble fields in the deep channel allows the achievement
of much higher electron energies with a laser of a given pulse energy.
Let us assume that the accelerating bubble field $E$ is completely
at our disposal and that the laser pulse energy is converted primarily
into that field. The laser pulse energy scales as $W_{{\rm L}}\propto R^{2}Ic\tau$,
where $I$ is the laser intensity. We omit dimensionless numerical
factors like $\pi$ etc. as we are interested in parametric dependencies
only. The numerical pre-factors can be obtained from gauging the scaling
laws against PIC simulations.

We can find the pulse depletion length $L_{{\rm d}}$ by comparing
the energy deposited in the wake field $W_{{\rm wake}}\propto E^{2}R^{2}L_{{\rm d}}$
and the initial pulse energy, $W_{{\rm wake}}=W_{{\rm L}}$:

\begin{equation}
L_{{\rm d}}\propto Ic\tau E^{-2}.\label{eq:depletion length}
\end{equation}
Here we assume a cylindrically symmetric bubble whose radius scales
together with the laser radius $R$.

The particle acceleration is limited either by the laser pulse depletion
\eqref{eq:depletion length}, or by electron dephasing. In the uniform
plasma case, it was the laser pulse depletion that limited acceleration
in the bubble regime. In the deep channel case of interest here, however,
the depletion length can be adjusted by properly choosing the accelerating
field $E$, as can be seen from \eqref{eq:depletion length}. The
maximum particle energy is achieved when the depletion length $L_{{\rm d}}$
equals the dephasing length $L_{{\rm A}}$. Otherwise, the maximum
energy gain is limited by the shortest of the two lengths and is lower.
The dephasing length scales as $L_{{\rm A}}\propto R\gamma_{{\rm L}}^{2}$,
where $\gamma_{{\rm L}}$ is the relativistic $\gamma-$factor associated
with the laser group velocity. In a deep plasma channel, $\gamma_{{\rm L}}$
is not influenced by the plasma and is defined solely by the laser
pulse radius so that we can write $\gamma_{{\rm L}}\propto k_{0}R$,
where $k_{0}=2\pi/\lambda_{0}$. For the dephasing length we obtain 

\begin{equation}
L_{{\rm A}}^{{\rm dc}}\propto k_{0}^{2}R^{3}\label{eq:dephasing}
\end{equation}
\begin{figure}
\includegraphics[width=0.9\columnwidth]{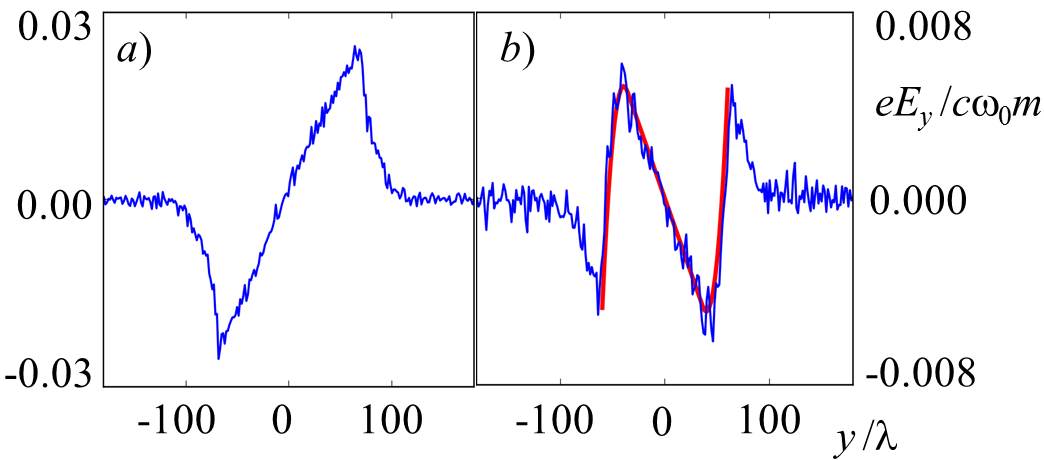}

\protect\caption{\label{fig:reversal} Transverse electric field of the bubble for
the laser radius $R=40\lambda_{0}$ and $a_{0}=10$. a) Homogeneous
plasma: the field is focusing. b) Channel: the field reverses its
sign in the channel walls; the smooth curve gives the analytic solution
for the field $E_{y}$.}
\end{figure}

We reuire the depletion length of Eq. \eqref{eq:depletion length}
to be equal the dephasing length \eqref{eq:dephasing} and obtain
for the bubble field $E\propto\sqrt{Ic\tau/k_{0}^{2}R^{3}}$ in the
deep plasma channel. This leads to the maximum energy gain scaling 

\begin{equation}
\mathcal{E}_{\max}^{{\rm dc}}\propto eL_{{\rm A}}E\propto e\sqrt{Ic\tau k_{0}^{2}R^{3}}=k_{0}\sqrt{mc^{2}W_{L}r_{e}R}\label{eq:max energy}
\end{equation}
The deep channel scaling \eqref{eq:max energy} differs dramatically
from the uniform plasma bubble scaling \eqref{eq:uniform}. While
the uniform plasma case requires the maximum pulse duration $\tau$,
in the deep channel, pulses with radius $R$ as large as possible
have an advantage. 

A deeper insight in the bubble field configuration provides the quasistatic
theory, which implies that the bubble slowly evolves in time and the
bubble fields depend on $\xi=t-z$ \cite{sprangle 1990,whittum,mora}.
The nonlinear theory of the bubble in homogeneous plasma \cite{scaling Lu}
can be generalized to transversally inhomogeneous plasma (details
be published elsewhere). It follows from the theory that the bubble
field not too close to the edge of the bubble is defined by the source
function $s(r_{b})\equiv(1/2r_{b}^{2})\int_{0}^{r_{b}}\rho_{ion}\left(r\right)rdr$: 

\begin{equation}
\mathbf{E}\approx\mathbf{e}_{r}\left[\frac{2r_{b}^{2}s(r)}{r}-rs(r_{b})\right]-\mathbf{e}_{z}2\xi s(r_{b}),\mathbf{\,\, B}\approx-rs(r_{b}).\label{E}
\end{equation}

For homogeneous plasma $\rho_{ion}\left(r\right)=1$ we recover known
expression for the bubble field \cite{Kost2004} $\mathbf{E}\approx\mathbf{e}_{r}r/4-\mathbf{e}_{z}\xi/2$
and $\mathbf{B}\approx-e_{\varphi}r/4$. For the plasma channel with
exponential profile discussed above the source function takes a form
$s(r)=|e|n_{0}\left(r_{0}^{2}-r^{2}+2\eta r_{ch}\delta\right)/\left(4r_{b}^{2}\right)$,\emph{
}for $r\geq r_{0}$ and $s(r)=0$ for $r<r_{0}$, where $\eta=\exp(r/r_{ch})\left(r-r_{ch}\right)+\exp(r_{0}/r_{ch})\left(r_{ch}-r_{0}\right)$.
The obtained solution says that inside the vacuum part of the plasma
channel $r<r_{0}$ the bubble field is purely electromagnetic $B_{\varphi}=E_{r}=2(r/\xi)E_{z}$.
In the channel walls, the ion field is added and can\emph{ reverse
the sign of the transversal electric field (see Fig.\ref{fig:reversal})
.} Another conclusion is that the plasma channel\emph{ reduces the
gradients of the accelerating and focusing forces} in comparision
with homogeneous plasma. 

\begin{figure}
\includegraphics[width=1\columnwidth]{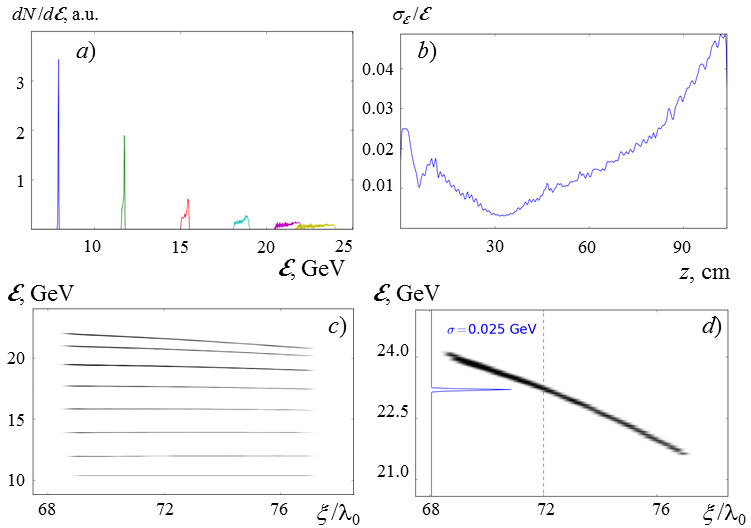}

\protect\caption{\label{fig:spectra} Simulation results for the case $R=32\,\mu$m
($R=40\lambda_{0}$). a) Energy spectra of the witness beam at different
acceleration stages. b) Evolution of relative r.m.s. spectral width
$\sigma_{\mathcal{E}{\rm i}}$. The homogenous acceleration lasts
for the first 30 cm and minumum spectral width of 0.3\% is reached.
Later, the bunch gains positive energy chirp due to dephasing and
the spectrum widens. c) Longitudinal phase space of the witness bunch
at different propagation distances. d) The final phase space of the
bunch zoomed. The local energy uncertainty is $10^{-3}$ and is limited
by numerical resolution.}
\end{figure}

To check the quasi-static theory and the energy scalings, we perform
a series of 3d PIC simulations using the code VLPL \cite{VLPL}. We
run the code in laboratory frame for simulations with the smallest
pulse radius. The laser pulse is circularly polarized and has the
envelope $a\left(t,{\bf r}\right)=a_{0}\exp\left(-r^{2}/R^{2}-t^{2}/\tau^{2}\right)$.
The envelope is cut to zero at $t=2\tau,\,\, r=2R$. We assume laser
wavelength $\lambda_{0}=800\,$nm. 

We observe very little or no self-injection in the bubble when an
empty on-axis channel is used. Thus, we inject an external co-propagating
witness electron bunch at the end of the bubble. It has the half-length
$l_{i}$, initial energy $\mathcal{E}_{{\rm i}}$, and the total charge
$N_{i}$. The simulation parameters are collected in Table \ref{fig:table}. 

\begin{figure}
\includegraphics[width=0.7\columnwidth]{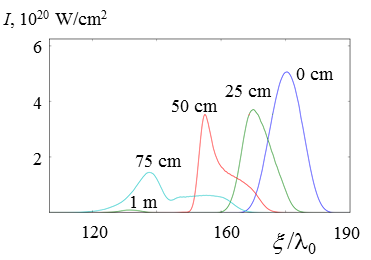}

\protect\caption{\label{fig:laser evolution} Nonlinear evolution of the laser pulse
for the simulation case $R=32\,\mu$m. No typical for bubble regime
pulse shortening or ``optical shock'' formation.}
\end{figure}

We selected an extremely short laser pulse with $\tau_{0}=4\,$fs,$R_{0}=8\,\mu$m,
and energy of $W_{{\rm L}}=2.2\,$J as the first member for the sequence
of three in our power of 2 scaling sequence for both pulse duration
and radius, as given in Table \ref{fig:table}. Although such short
and energetic pulses do not exist yet, projects are under way to achieve
similar parameters. According to the similarity rule, we scale simultaneously
the laser pulse radius and duration with the same factor $\alpha=2$
from stage $n$ to stage $n+1$: $R_{(n+1)}=\alpha R_{(n)}$ and $\tau_{(n+1)}=\alpha\tau_{(n)}$
so that the aspect ratio $\Pi=R/c\tau\approx6.7$ remains fixed, i.e.
the pulses are pancake-like. This allows us to reproduce the wake
field (the bubble) \textit{exactly} without additional search for
the optimal plasma channel parameters. The other parameters scale
as $R_{{\rm ch(n+1)}}=\alpha R_{{\rm ch(n)}}$, $n_{e(n+1)}=\alpha^{-2}n_{e(n)}$,
$E_{(n+1)}=\alpha^{-1}E_{(n)}$, $L_{{\rm A(n+1)}}=\alpha^{3}L_{{\rm A}(n)}$,
and the particle energy scales as $\mathcal{E}_{\max(n+1)}=\alpha^{2}\mathcal{E}_{\max(n)}$.
The three scaled cases are digested in Table \ref{fig:table}. The
most energetic pulse with $\tau_{2}=16\,$fs, radius $R_{2}=32\,\mu$m
and energy of $W_{{\rm L}}=141\,$J would correspond to the planned
Apollon laser \cite{apollon}. The acceleration lengths range from
$L_{{\rm A}}=2\,$cm for the shortest laser, to $L_{{\rm A}}=1\,$m
for the largest laser pulse. These distances can be simulated only
using the Lorentz boost \cite{vey}. 

The bubble generated by the largest laser pulse radius $R=32\,\mu$m
is shown in Fig.\ref{fig:bubble}. The accelerating field, Fig.\ref{fig:bubble}(a),
is transversely uniform. It allows for mono-energetic acceleration
of wide electron bunches. This transverse field uniformity is also
observed in homogeneous plasmas. The on-axis profile of the accelerating
field in the channel, Fig.\ref{fig:bubble}(b), however, differs from
that in the uniform plasma case. There is a region of flat accelerating
field at the very back of the bubble. This region can be used to accelerate
reasonably long witness bunches mono-energetically. 

Fig.\ref{fig:reversal} shows the bubble transverse fields in the
bubble corresponding to the case $R=40\lambda_{0}$ and $a_{0}=10$.
We compare predictions of the quasi-static model with the numerical
simulations. In the bubble center, $dE_{y}/dy\approx0.08$, $dE_{z}/d\xi\approx0.15\approx2(dE_{y}/dy)$
in the simulation and $dE_{y}/dy\approx0.08$, $dE_{z}/d\xi\approx0.16=2(dE_{y}/dy)$
in the models that is in a very good agreement. Fig.\ref{fig:reversal}(b)
clearly shows the field reversal in the channel.

To check the acceleration, we inject a witness bunch that occupies
about 10\% of the bubble length. The energy spectra of the accelerated
bunch are shown in Fig.\ref{fig:spectra}(a) at different times. The
bunch stays very monoenergetic for the first 30 cm of acceleration,
because it was injected in the flat accelerating field part of the
bubble. The relative r.m.s. energy spread $\sigma_{\mathcal{E}}/\mathcal{E}$,
Fig.\ref{fig:spectra}(b), of the bunch is merely 0.3\% when it gains
7.5 GeV energy. This is much better than the ratio of the bunch length
to the bubble length that is 10\%. Later, the bunch slowly leaves
this flat $E-$field region and advances into the region with linearly
growing $E_{z}-$field. Finally, it gains a positive energy chirp
as seen in the bunch longitudinal phase space, Fig.\ref{fig:spectra}(c). 

However, the quality of acceleration is best understood when we zoom
in at the longitudinal phase space of the witness bunch, Fig.\ref{fig:spectra}(d).
The r.m.s relative width of the witness bunch energy uncertainty taken
at a particular longitudinal position is merely $10^{-3}$ and is
probably defined by numerical resolution of our code. This very narrow
energy spread is due to (i) the transverse uniformity of accelerating
field in the bubble and (ii) the lack of a betatron resonance in the
hollow channel.

Fig. \ref{fig:laser evolution} shows the nonlinear evolution of the
laser pulse. Very different from the uniform plasma case, we observe
no laser pulse shortening and no optical shock formation. The channel
parameters were chosen to balance the dephasing length and the laser
depletion length: $L_{{\rm A}}=L_{{\rm D}}=1\,$m.

In conclusion, we have shown that electron acceleration in deep plasma
channels is scalable and the energy scalings favor ultra-short pancake-like
laser pulses. The accelerating field has a nearly flat field region
at the rear, where the accelerating field depends only slightly on
the longitudinal coordinate. This allows for nearly monoenergetic
acceleration of a witness bunch. 

This work has been supported by the Deutsche Forschungsgemeinschaft
via GRK 1203 and SFB TR 18, by EU FP7 project EUCARD-2 and by the
Government of the Russian Federation (Project No. 14.B25.31.0008)
and by the Russian Foundation for Basic Research (Grants No. 13-02-00886,
13-02-97025).


\begin{thebibliography}{References}
\bibitem{esarey} E. Esarey, C. B. Schroeder, and W. P. Leemans, Rev.
Mod. Phys. \textbf{81}, 1229 (2009) 

\bibitem{Joshi} C. Joshi and A. Caldwell, \textquotedblleft Plasma
Accelerators,\textquotedblright{} Handbook for Elementary Particle
Physics, Volume III, edited by S. Myers and H. Schopper, Heidelberg,
Springer (2013), 12-1.

\bibitem{bubble} A. Pukhov, J. Meyer-ter-Vehn, Applied Physics\textbf{
B 74}, 355 (2002)

\bibitem{Kost2004}I. Kostyukov, A. Pukhov, S. Kiselev, Phys. Plasmas
\textbf{\textcolor{black}{11}}, 5256 (2004)

\bibitem{Malka 2013}V. Malka \textit{Laser Plasma Accelerators,}
in Laser-Plasma Interactions and Applications, Scottish Graduate Series,
pp 281-301 (2013)

\bibitem{scaling Lu}W. Lu et al. PRSTAB \textbf{10}, 061301 (2007)

\bibitem{BlowoutTheory Lu} W. Lu et al., Phys. Rev. Lett. \textbf{96},
165002 (2006).

\bibitem{similarity}S. Gordienko, A. Pukhov Physics of Plasmas \textbf{12},
043109 (2004).

\bibitem{scaling PG}A. Pukhov, S. Gordienko, Phil Tr. R. Soc. A,
\textbf{364}, 623 (2006).

\bibitem{Jansen2014}O. Jansen, T. Tückmantel, and A. Pukhov, Eur.
Phys. J. Special Topics \textbf{223}, 1017\textendash 1030 (2014)

\bibitem{DLA}A Pukhov, ZM Sheng, J Meyer-ter-Vehn, Physics of Plasmas
\textbf{6}, 2847 (1999)

\bibitem{Dino Nature}S. Cipiccia et al Nature Physics \textbf{7,}
867 (2011) 

\bibitem{DLA Mori}J L Shaw et al., arXiv:1404.4926 (2014)

\bibitem{Schroeder 2013}C. B. Schroeder, E. Esarey, C. Benedetti,
W. Leemans Phys.Plasmas \textbf{20}, 080701 (2013) 

\bibitem{Chiou}T. C. Chiou et. al., Physics of Plasmas \textbf{2},
310 (1995) 

\bibitem{Faure 2005 shortening}J. Faure et al., Phys. Rev. Lett.
\textbf{95}, 205003 (2005).

\bibitem{ziegler 2009}Y. Katzir et al. Appl. Phys. Lett. \textbf{95},
031101 (2009).

\bibitem{sprangle 1990}P. Sprangle, E. Esarey, and A. Ting, Phys.
Rev. Lett. \textbf{64}, 2011 (1990)

\bibitem{whittum}D. H. Whittum, Phys. Plasmas, \textbf{4}, 1154 (1997)

\bibitem{mora}P. Mora and Th. M. Antonsen Jr., Phys. Plasmas \textbf{4},
(1997)

\bibitem{VLPL}A. Pukhov Journal of Plasma Physics \textbf{61}, 425
(1999).

\bibitem{apollon}F. Giambruno et al., Appl. Opt. \textbf{17}, 2617
(2011).

\bibitem{vey} J.-L. Vay, C. G. R. Geddes, E. Cormier-Michel, and
D. P. Grote, Phys. Plasmas \textbf{18}, 030701 (2011)\end{thebibliography}
\end{document}